\documentclass[showpacs,prl,twocolumn]{revtex4}

\usepackage{epsfig,amsmath}

\usepackage{subfigure}

\usepackage{graphicx}

\usepackage{dcolumn}

\usepackage{stmaryrd}

\usepackage{mathrsfs}

\usepackage{pifont}

\usepackage{amsthm}

\usepackage{amssymb}

\usepackage{bm}

\usepackage{latexsym}

\usepackage{hyperref}

\usepackage{color}

\usepackage{amsmath}

\usepackage{appendix}

\newcommand{\beq}{\begin{equation}}

\newcommand{\eeq}{\end{equation}}

\newcommand{\beqa}{\begin{eqnarray}}

\newcommand{\eeqa}{\end{eqnarray}}

\begin{document}

\title{Non-Markovian environment induced anomaly in steady state quantum coherence}


\begin{abstract}

	Environment induced steady state quantum coherence (SSQC) is a captivating phenomenon that challenges conventional understandings of decoherence. In this letter, we delve into the foundational aspects of environment-induced SSQC, shedding light on its emergence within the framework of system-bath interactions. Starting from a microscopic system-bath coupled model, we investigate the dependence of SSQC on environmental memory effects, bath temperature, system-bath coupling strength, and squeezing parameters. Our findings reveal that the environment not only acts as a generator but also as a disruptor of SSQC. A peak will exist for a non-Markovian bath, which is a result of competition between these two mechanisms. Interestingly, the peak disappears in Markovian case. Additionally, we observe that the generated SSQC can be further amplified through environment squeezing.	

\end{abstract}

\author{Arapat Ablimit$^{1}$, Zhao-Ming Wang$^{1}$\footnote{wangzhaoming@ouc.edu.cn}, Feng-Hua Ren$^{2}$, Paul Brumer$^{3}$, and Lian-Ao Wu$^{4,5,6}$}

\affiliation{$^{1}$ College of Physics and Optoelectronic Engineering, Ocean University of China, Qingdao 266100, China \\
	$^{2}$ School of Information and Control Engineering, Qingdao University of Technology, Qingdao 266520, China\\
	$^{3}$Chemical Physics Theory Group, Department of Chemistry, and Center for Quantum Information and Quantum Control, University of Toronto, Toronto, Ontario, Canada M5S 3H6\\
	$^{4}$ Department of Physics, University of the Basque Country UPV/EHU, 48080 Bilbao, Spain\\
	$^{5}$IKERBASQUE Basque Foundation for Science, 48013 Bilbao, Spain\\
	$^{6}$ EHU Quantum Center, University of the Basque Country UPV/EHU, Leioa, Biscay 48940, Spain}

\maketitle

\emph{Introduction.}\label{1}
Quantum coherence is a foundational element in quantum information processing, responsible for the advantages that quantum tasks have over classical ones \cite{Nielsen,Cina,SKoyu}. It also plays an important role in emergent fields such as nanoscale devices thermodynamics \cite{Tajima} and quantum biology \cite{Kominis}. In the realms of quantum computation and information processing, the preservation of steady state quantum coherence  (SSQC) over extended periods is typically required. Nevertheless, the maintenance of coherence in quantum systems proves challenging due to the decoherence and dissipation \cite{Eric1,Eric2} induced by system-bath interactions \cite{Schlosshauer}. This presents a substantial hurdle to the progress of quantum technology. Consequently, extensive effort has been devoted to developing strategies that can counteract or alleviate the adverse impacts of the environment \cite{Shor, Lidar, Rabitz, wanghybird}. Initially, it was believed that the environment would always induce decoherence in the system. However, researchers have discovered that the environment can actually play a constructive role \cite{MBPlenio,ADodin, SFHuelga, CGuarnieri, Cui, Reppert, ZWang}. Additionally, long-lasting coherence has also been observed in the excitation energy transport of certain photosynthetic pigment-protein complexes \cite{Mohseni, PBrumer18} under pulsed laser excitation.

 When considering the system-bath interaction, the dynamics of the open quantum system can be described by the master equation \cite{Piilo09}. The well-known Lindblad master equation is commonly used for memoryless Markovian processes \cite{Lindblad, VGorini}. However, when the memory effect can not be neglected, a non-Markovian description is indispensable \cite{Vega}. Various methods have been developed to address non-Markovian dynamics, and quantum state diffusion (QSD) is an effective one \cite{Strunz99,Gisin}. The non-Markovianity has shown diverse effects, including  the storage of quantum information \cite{Vega,wanghybird}, heat control \cite{wangheat}, and adiabatic speedup \cite{wang2018}.

 Recently, environment induced quantum coherence \cite{Pozzoboma} has been studied.  The SSQC has been observed in a two-level system within a non-Markovian environment\cite{AAsk, SFHuelga}. Environment induced decoherence \cite{JPPaz,WMZhang} and coherence \cite{MBPlenio,Cui,Ancheyta} has been extensively studied, motivating the question as to whether it it possible for the environment to play both a coherence generative and coherence destructive role simultaneously? In this letter we examine this issue by using a system-bath coupling model, where $N$ non-interacting qubits are immersed in a collective environment. Initially, no coherence exists among the qubits. Our investigation delves into the impact of the environment on SSQC. We observe that SSQC initially rises and then declines as the system-bath interaction strength, environmental Markovianity, and temperature increase. This indicates that coherence generation dominates in a weak environment, while the impact of destructiveness becomes more significant in a relatively stronger environment. Furthermore, we demonstrate the SSQC can be further enhanced by squeezing the environment in a suitable range of the squeezing parameters.

\emph{Model and method.}\label{2} 
For a quantum system embedded in an environment, the total Hamiltonian can be written as $H_{\text{tot}} = H_{\text{s}} + H_{\text{b}} + H_{\text{int}}$, where $H_{\text{s}}$ and $H_{\text{b}}$ are the Hamiltonians of the system and bath, respectively, and $H_{\text{int}}$ describes the system-bath interactions. In this letter we consider the linear system-bath coupling, which has been used to characterize the dynamics of physical systems such as quantum dots in an electron bath \cite{PBethke,MChen}, and atoms in electromagnetic field \cite{wang2018,zhao2}. The Hamiltonians $H_{\text{b}} = \sum_{k} \omega_{k}b_{k}^{\dagger} b_{k}$ and $H_{\text{int}} = \sum_{k} \left( f_{k}^{\ast} L^{\dagger} b_{k} + f_{k} L b_{k}^{\dagger} \right)$, where $b_{k}^{\dagger}$ ($b_{k}$) denotes the creation (annihilation) operator of the $k$th mode in the bath. $L$ is the Lindblad operator describing the system operator with a coupling strength $f_{k}$. We let $\hbar = k_B = 1$ throughout.

Suppose initially the whole system is in a direct product state $\rho_{tot}(0)=\rho_{s}(0)\otimes\rho_{\text{b}}(0)$, where $\rho_{s}(0)$ and $\rho_{\text{b}}(0)$ are system and environment initial state, respectively. The system is in a pure state $\left|\psi(0)\right\rangle $ and the bath is at thermal equilibrium state with the density operator $\rho_{\text{b}}(0)=e^{-\beta  H_{\text{b}}}/Z$. Here $Z=Tr(e^{-\beta H_{\text{b}}})$  is the partition function and $\beta=1/T$.
In the following we use the non-Markovian master equation \cite{Gisin} derived from the QSD method to calculate the dynamics of the system. For bosonic thermal bath, the bath creation and annihilation operators obeys the commutation relation $[b_{i},b_{j}^{\dagger}]=\delta_{i,j}$, and the general non-Markovian master equation reads \cite{yu2004,wangheat} 

\begin{eqnarray}
	\frac{\partial}{\partial t}\rho_{s}&=&-i\left[H_{s},\rho_{s}\right]+[L,\rho_{s}\overline{O}^{\dagger}(t)]-[L^\dagger,\overline{O}(t)\rho_{s}]\nonumber\\&\;&+[L^\dagger,\rho_{s}\overline{Q}^{\dagger}(t)]-[L,\overline{Q}(t)\rho_{s}].
	\label{eq4}
\end{eqnarray}
Note that in this letter we consider weak system-bath interaction, the $\overline{O}(t)$, $\overline{Q}(t)$ operators have been approximated by noise-independent \cite{Strunz99,Yu2000} in above equation. $\overline{O}(t)=\int_{0}^{t}ds\alpha(t,s)O(t,s)$, $\overline{Q}(t)=\int_{0}^{t}ds\eta(t,s)Q(t,s)$ and $\alpha(t,s)$, $\eta(t,s)$ are the bath correlation functions. At thermal bath, these correlation functions given by $\alpha(t,s)=\int d\omega J(\omega)(\overline{n}_{k}+1)e^{-i \omega_{k} (t-s)}$, $\eta(t,s)=\int d\omega J(\omega)\overline{n}_{k}e^{i \omega_{k} (t-s)}$ , where $\overline{n}_k=\frac{1}{e^{\omega_k/T}-1}$ is the mean thermal occupation number of quanta in mode $\omega_k$, and $J(\omega)$ is the bath spectral density.  The $O$, $Q$ operators are defined by an ansatz with  initial conditions $O(t,s=t)=L$, $Q(t,s=t)=L^\dagger$.  In general, the two operators contain the environmental stochastic noises $z^{\ast}_t,w^{\ast}_t$, and can be  exactly obtained for some simple models \cite{Gisin}. But for most of the models obtaining the noise-dependent exact $O$, $Q$ operators are extremely difficult. The $\overline{O}(t)$, $\overline{Q}(t)$ operators satisfy

\begin{eqnarray}
	\frac{\partial}{\partial t}\overline{O}(t)&=& \alpha(0,0)L-(iw_{0}+\gamma)\overline{O}(t)-[\mathcal A,\overline{O}(t)],
	\label{eq10}\\
		\frac{\partial}{\partial t}\overline{Q}(t)&=& \eta(0,0)L^\dagger+(iw_{0}-\gamma)\overline{Q}(t)-[\mathcal A,\overline{Q}(t)],
	\label{eq11}
\end{eqnarray} 
where $\mathcal A=iH_{s}+L^\dagger\overline{O}(t)+L\overline{Q}(t)$. We use the Ohmic type spectrum with a Lorentz–Drude cutoff \cite{Meier,Ritschel} as an example, it is defined as $J(\omega)=\frac{\Gamma}{\pi} \frac{\omega\gamma^{2}}{\gamma^{2}+\left(\omega_{0}-\omega\right)^{2}}$. The parameters $\Gamma$ and $\gamma$ characterize the system-bath coupling strength and bandwidth, respectively. When $\gamma\rightarrow0$, the environment has a strong non-Markovianity, while $\gamma\rightarrow\infty$ corresponds to the white noise case i.e., Markov limit. Here we take Hermitian Lindblad operator $L=\sum_{i} \sigma_{i}^x$ throughout, which corresponds to the spin-boson interaction.

At first, assume the environment is initially at vacuum state $\left|0\right\rangle$. We define it as environment with initial vacuum state (EIVS). In the following, we incorporate the solution of Eqs.~(\ref{eq10})-(\ref{eq11}) in the non-Markovian master equation (Eq.~(\ref{eq4})) for the numerical calculation. In the Markovian limit, the dynamics are governed by the Markovian Lindblad master equation \cite{wang2020,Zhao Fermion}, 
\begin{eqnarray}
	\frac{\partial}{\partial t}\rho_{s}&=&-i\left[H_{s},\rho_{s}\right]+\frac{\Gamma T}{2}[\left(2L\rho_{s}L^{\dagger}-L^{\dagger}L\rho_{s}-\rho_{s}L^{\dagger}L \right) \nonumber\\&\;&
	 +\left(2L^{\dagger}\rho_{s}L-LL^{\dagger}\rho_{s}-\rho_{s}LL^{\dagger}\right)].
	\label{eq32}
\end{eqnarray}

For the second case, the initial states of the bath are at squeezed state $s(\xi)\left|0\right\rangle$,  $s(\xi)$is a unitary squeezing operator. We  denote it as the environment with initial squeezed state (EISS). Compared to the first case, the quantum fluctuations of one quadrature component of a squeezed-state can be smaller than the symmetry limit $\hbar /2$. This property has been widely used in the fields of quantum metrology \cite{Iwasawa,Peano,MATaylor} and quantum control \cite{Jabri,Arapat zero}. In the following, we consider a symmetric two-mode squeezed bath, it can be generated by pumping a lumped-element Josephson parametric amplifier in superconducting circuit \cite{Murch}. In this case
\begin{eqnarray}
	s(\xi)=e^{r(\xi^{\ast}b_kb_\lambda-\xi b_k^{\dagger}b_\lambda^{\dagger})}, 
	\label{eq323}
\end{eqnarray}
where $\xi=e^{i\theta}$. $r$ is the squeezing strength, and $\theta$ denotes the squeezing direction. In fact, the two mode squeezing generates coherence between the two oscillators \cite{GAdesso}. For the dynamics of the system in the squeezing environment, see the Appendix.

We consider an uncoupled $N$ qubits model, with the system Hamiltonian $H_{s}= \sum_{i=1}^{N} \omega_i\sigma_{i}^z$, where $\sigma_{i}^z$ is the Pauli matrices of a general two level system with atomic transition frequency $\omega_{i}$.  For simplicity, in the following study we take $\omega_i$=$\omega_{s}$ for all $i$.

\emph{ Environment induced steady state quantum coherence.}\label{3}
 \begin{figure}

 	(a)

 	\centerline{\includegraphics[width=1\columnwidth,height=0.77\columnwidth]{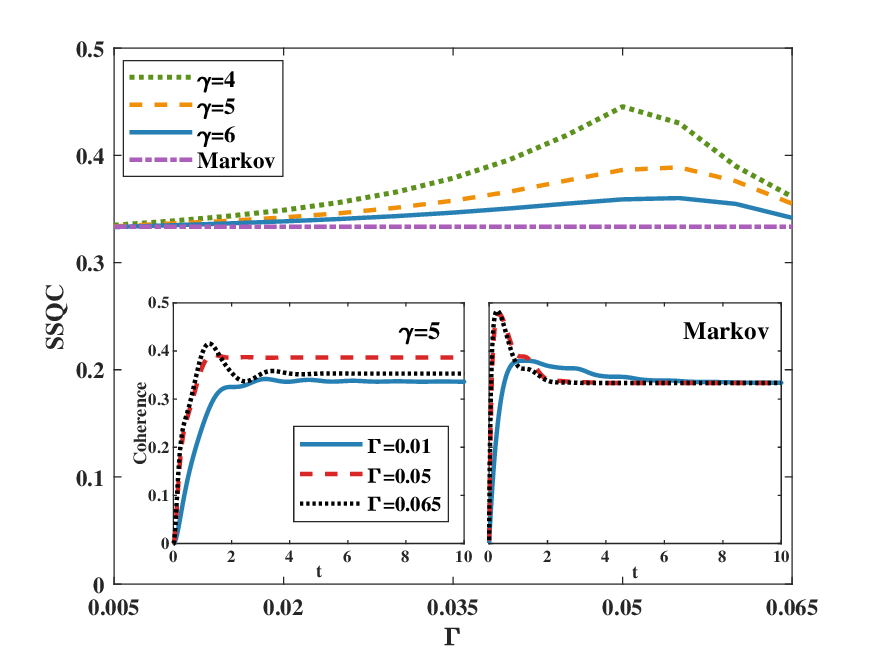}}

 	(b)

 	\centerline{\includegraphics[width=1\columnwidth,height=0.77\columnwidth]{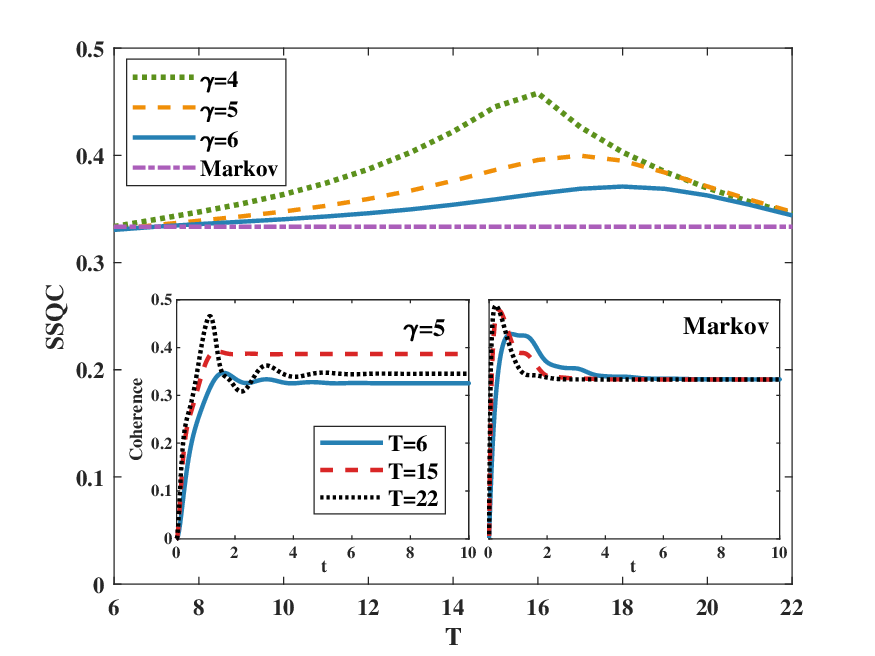}}

 	(c)

    \centerline{\includegraphics[width=1\columnwidth,height=0.77\columnwidth]{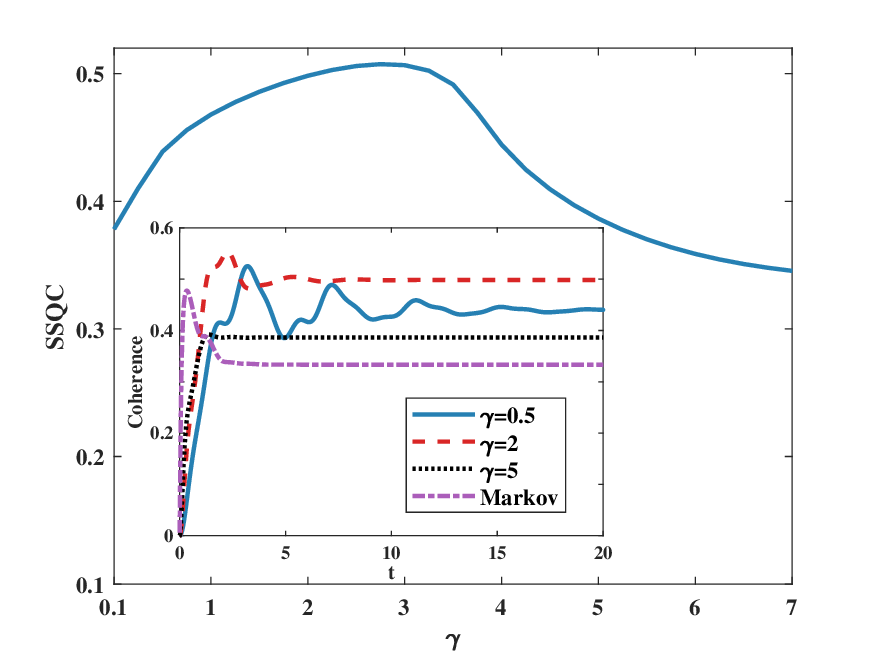}}

 	\caption{(Color on line) The SSQC dynamics versus environmental parameters (a) $\Gamma$, $T=15$; (b) $T$, $\Gamma=0.05$; (c) $\gamma$, $\Gamma=0.05$, $T=15$. Here the SSQC value is taken in the steady state limit. Inset given the time evolution of quantum coherence for different parameters (a) $\Gamma$, with $\gamma=5$ and Markov limit; (b) T, with $\gamma=5$ and Markov limit; (C) $\gamma$ . Other parameters are set as $\omega_0=\omega_s=1$, and $N=2$.}
 	\label{fig:1}	 
 \end{figure}
The coherence measure we used in this letter is based on the $l_{1}$ norm \cite{Streltsov,Baumgratz2014}, which is defined as the sum over the absolute value of the off-diagonal elements i.e.,  $C\left(t\right)=\sum_{i\neq j}\left|\rho(t)_{i,j}\right|$ with $\left\langle i\right|\rho(t)\left|j\right\rangle$. Here $i$ and $j$ label individual energy eigenstate in each of the two level systems. Below, on axis labels, ``SSQC'' denotes $C(t)$ in the steady state limit. Note that the size of populations, i.e., the typical value of $\rho_{ii}$ can also help us to understand the relative size of the coherence. Taking $N=2$ as an example, we plot the SSQC values for different system-bath interaction intensity $\Gamma$, temperature $T$, and environmental memory capacity $\gamma$ in Fig.~\ref{fig:1} (a-c), respectively. The other environmental parameters are set as $T=15$ (Fig.~\ref{fig:1} (a)), $\Gamma=0.05$ (Fig.~\ref{fig:1} (b)), and $\Gamma=0.05$, $T=15$ (Fig.~\ref{fig:1} (c)). From Fig.~\ref{fig:1} we can see a very interesting phenomena that SSQC first increases then decreases with increasing $\Gamma$, $T$ or $\gamma$. It has a peak for certain parameters. On the one hand, the collective environment introduces the interaction between the two qubit, thereby the induced coherence becomes larger when the environment becomes stronger (bigger $\Gamma$, $T$ or $\gamma$). On the other hand, the environment also plays adverse role on the coherence. So there is a competition between these two mechanism. The former (later) dominates in a weaker (stronger) environment. The peak indicates this transition. In other words, for two given environmental parameters, we can always find the third parameter that corresponds to the maximum coherence. In the inset of Fig.~\ref{fig:1}(a-c), we plot the time evolution of the quantum coherence for different environmental parameters. From the inset of Fig.~\ref{fig:1} (a),(b), for a strong environment, the coherence first increases quickly, reaches a peak, then decreases. On the contrary, for a weak environment, the coherence increases gradually without oscillations. However, in the inset of Fig.~\ref{fig:1}~(c), the similar behavior is not observed since $1/\gamma$ represents the memory time. Smaller $\gamma$ corresponds to stronger non-Markovinity. In this case, the information and energy exchanges frequently and correspondingly the oscillations exist, which has been observed in numerous references \cite{zhao2,ARivas,Harikrishnan}. Longer time is required to settle down to the SSQC in a more non-Markovian environment. Nevertheless, the coherence with a peak still holds. In addition, SSQC can even be induced in a Markovian environment, which is different from dephasing ($L=\sum_{i} \sigma_i^{z}$) and dissipation ($L=\sum_{i} \sigma_{i}^-$) coupling channels \cite{zhao2,SFHuelga}. To compare the non-Markovian and Markovian cases on the SSQC, in Fig.~\ref{fig:1} (a)-(b) we plot the SSQC versus $\Gamma$ and $T$ for different $\gamma$. Clearly in Markovian case, the SSQC is a constant, i.e., the SSQC of the system does not depend on the environmental temperature or strength. Also in the inset, $\Gamma$ or $T$ only affects the relaxation time to the steady state. The peak of the SSQC arises as an anomalous feature associated with the non-Markovianity of the bath. It is higher for a non-Markvoian bath and moves toward low temperature or weak coupling.

\begin{figure}

	\centerline{\includegraphics[width=1\columnwidth,height=0.9\columnwidth]{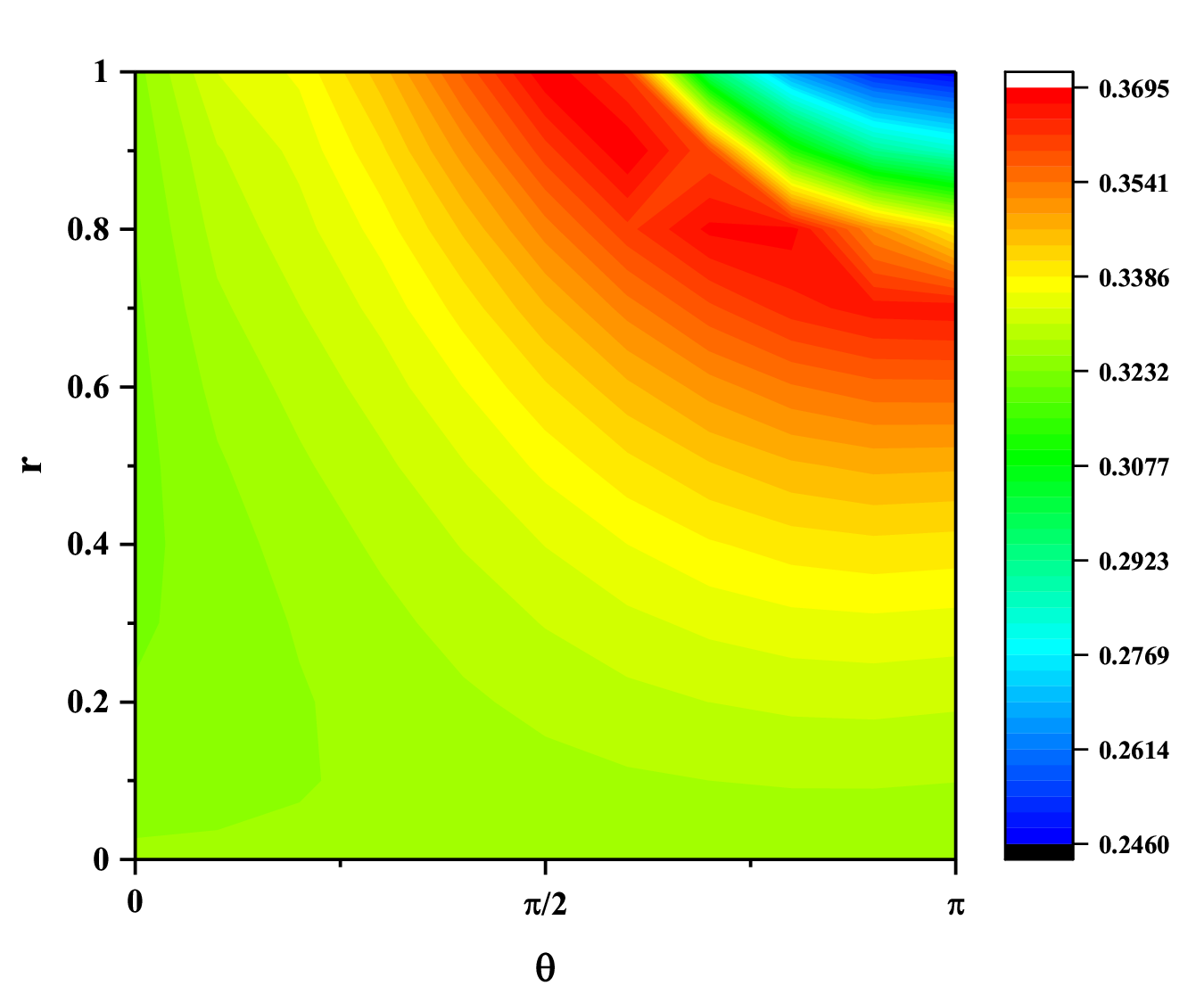}}

	\caption{(Color on line) The effect of reservoir squeezed parameters on SSQC. Other parameters are set as $\gamma=3$, $\Gamma=0.04$, $T=8$, $\omega_{0}=\omega_{s}=1$, and $N=2$.}

	\label{fig:2}	 

\end{figure}

\begin{figure}

	(a)

	\centerline{\includegraphics[width=1\columnwidth,height=0.8\columnwidth]{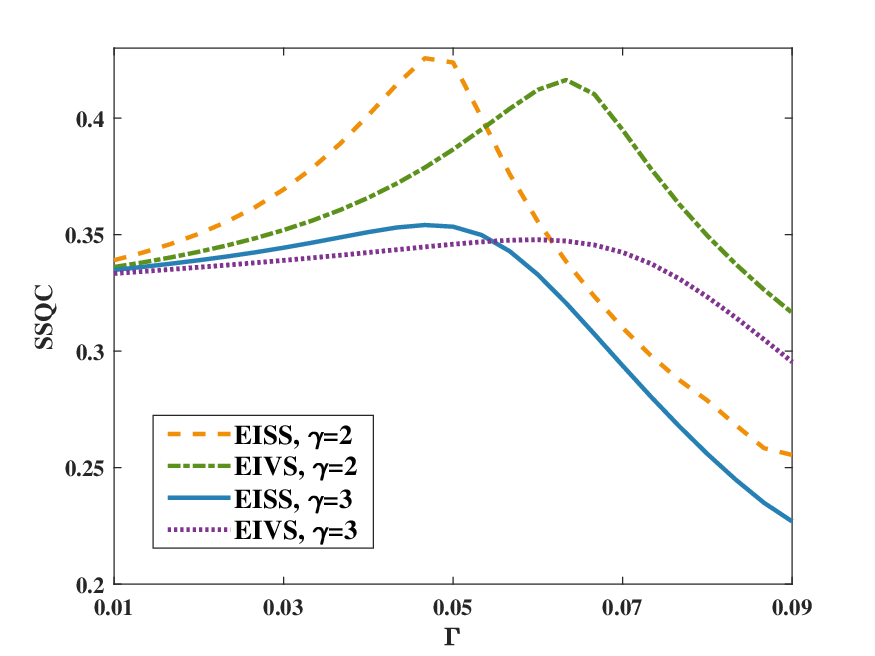}}

	(b)

	\centerline{\includegraphics[width=1\columnwidth,height=0.9\columnwidth]{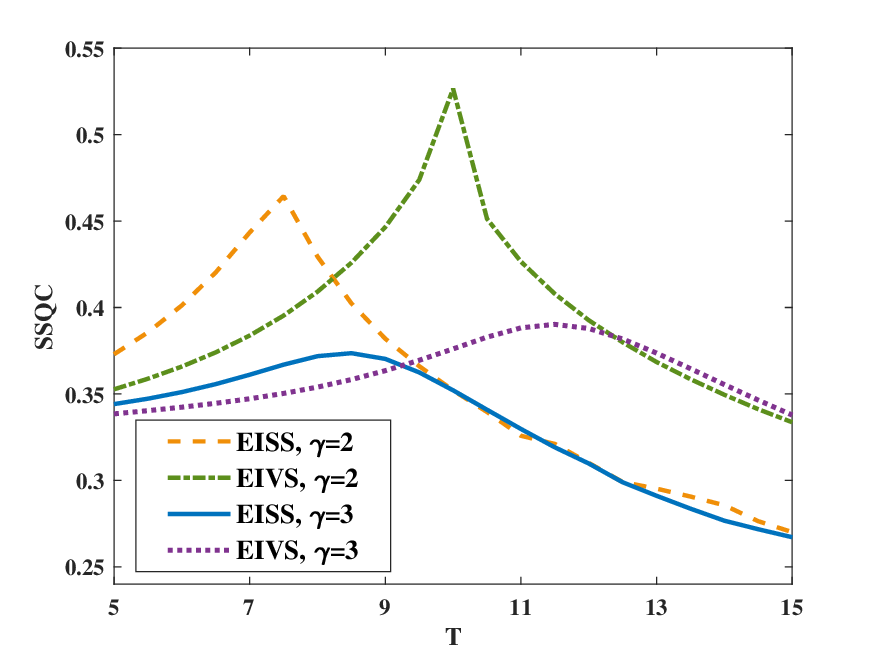}}

	\caption{(Color on line) The SSQC as functions of environmental parameters (a) $\Gamma$, $T=6$; (b) $T$, $\Gamma=0.04$ for different environmental initial state. Other parameters are taken as $r=0.4$, $\theta=\pi/2$, $\omega_{0}=\omega_{s}=0.5$, $N=2	$.}

	\label{fig:3}	 

\end{figure}

With the development of reservoir engineering technology, different initial states have been used to construct non-Markovian environments such as, initial vacuum state and squeezed state \cite{BLiu,Murch}. Among them, squeezed state acts as a special electromagnetic field state that corresponds to lower noise fluctuations than ordinary vacuum levels. This property has been widely used in the fields of quantum metrology \cite{Iwasawa,Peano,MATaylor} and quantum control \cite{Jabri,Arapat zero}. The fidelity of the adiabaticity \cite{Arapat zero} or quantum state transmission \cite{SPRA} can be enhanced by squeezing the environment. Can the generated coherence be further enhanced by squeezing? 
In Fig.~\ref{fig:2} we plot the effects of the squeezing on the SSQC (more details for numerical simulations see appendix). The environmental parameters are set as $\gamma=3$, $\Gamma=0.04$, $T=8$, $\omega_0=\omega_s=1$. It clearly shows that there is a parameter region that the SSQC reaches its maximum.
For the strongest squeezing $r=1$, SSQC first increases then decreases, reaches its maximum as $\theta$ from $0$ to $\pi$ (the corresponding direction from $p$ to $x$). A similar phenomena occurs when the bath is strongly squeezed in the x-quadrature ($\theta=\pi$) as $r$ increases. This is due to the  $x$-quadrature squeezing can strengthen the system-bath coupling, while the $p$-quadrature squeezing weaken it \cite{Link}.

Does the observed results in Fig.~\ref{fig:1} still hold in the case of squeezing? In Fig.~\ref{fig:3} we plot the SSQC as functions of different environmental parameters in the EISS. The squeezing parameters $r=0.4, \theta=\pi/2$. We also plot the EIVS case for the comparison. At first, the same behavior that SSQC first increases then decreases occurs in the EISS case. Secondly, for certain squeezing parameters, from Fig.~\ref{fig:3}(a) the peak of SSQC in the EISS is higher than in the EIVS, demonstrating the enhancement of the SSQC by squeezing. However, Fig.~\ref{fig:3}(b) shows that the SSQC is even lower in the EISS case for some certain squeezing parameters. At last, the SSQC is lower for a more Markovian environment in both EISS and EIVS cases. It is interesting that for some certain parameters, a higher SSQC in a more Markovian EISS can be obtained compared with the EIVS, such as in the right part of Fig.~\ref{fig:3}(a) (EIVS $\gamma=2$ and EISS $\gamma=3$).

\emph{Conclusions.}\label{4}
We use the QSD equation approach to investigate non-Markovianity induced SSQC. The system consists of a collection of uncoupled qubits that are linearly connected to a collective bath composed of harmonic oscillators with symmetric spin-boson model. Our results indicate that this collective bath can effectively generate SSQC.

We find that SSQC first increases then decreases with increasing $\gamma$, $\Gamma$ and $T$, which shows the competition between the generation and destruction of the coherence of the environment. A peak coherence always exists for a non-Markovian environment. 
Furthermore, we show that the SSQC can be further enhanced by squeezing the reservoir. The maximum corresponds to the strong squeezing in $x$-quaranture. Our results demonstrate the existence of maximum steady state coherence for certain environmental parameters.

\emph{Acknowledgements.} This work is supported by the Natural Science Foundation of Shandong Province (Grant No. ZR2021LLZ004). The contribution of P.B. is based upon work supported by the Air Force Office of Scientific Research under award FA9550-20-1-0354. L.-A.W. is supported by the Basque Country Government (Grant No. IT1470- 22) and Grant No. PGC2018-101355-B-I00 funded by MCIN/AEI/10.13039/501100011033.

\begin{widetext}
	
	\begin{appendix}
		
		\section*{ Appendix: A. Derivations of the non-Markovian master equation}
		
		\setcounter{equation}{0}
		
		\renewcommand\theequation{A.\arabic{equation}}
		
		In this appendix, we give a detailed derivation of the non-Markovian master equation. The wave function of the system can be obtained by projecting the coherent state of the bath modes onto the total wave function of the total system. The pure state for the system satisfies \cite{Strunz99,yu2004}	
		
		\begin{eqnarray}
			\frac{\partial}{\partial t}\left|\psi(t,z^{\ast}_t,w^{\ast}_t)\right\rangle &=&
			[-iH_{s}+Lz^{\ast}_t+L^\dagger w^{\ast}_t-L^\dagger\overline{O}(t,z^{\ast}_t,w^{\ast}_t)
			-L\overline{Q}(t,z^{\ast}_t,w^{\ast}_t)]\left|\psi(t,z^{\ast}_t,w^{\ast}_t)\right\rangle,
		\end{eqnarray}
		where $z^{\ast}_t,w^{\ast}_t$ are colored, complex stochastic environmental noises. The $\overline{O}$, $\overline{Q}$ operators contain the environmental memory effect, and are defined by
		
		\begin{eqnarray}
			\overline{O}(t,z^{\ast}_t,w^{\ast}_t)&=&\int_{0}^{t}ds\alpha(t,s)O(t,s,z^{\ast}_t,w^{\ast}_t),\\
			\overline{Q}(t,z^{\ast}_t,w^{\ast}_t)&=&\int_{0}^{t}ds\eta(t,s)Q(t,s,z^{\ast}_t,w^{\ast}_t),
			\label{eq7}
		\end{eqnarray}
		with functional derivative ansatz
		
		\begin{eqnarray}
			O(t,s,z^{\ast}_t,w^{\ast}_t)\left|\psi(t,z^{\ast}_t,w^{\ast}_t)\right\rangle =\frac{\delta}{\delta z^{*}_s} \left|\psi(t,z^{\ast}_t,w^{\ast}_t)\right\rangle,\\
			Q(t,s,z^{\ast}_t,w^{\ast}_t)\left|\psi(t,z^{\ast}_t,w^{\ast}_t)\right\rangle =\frac{\delta}{\delta w^{*}_s} \left|\psi(t,z^{\ast}_t,w^{\ast}_t)\right\rangle.
		\end{eqnarray}
		The $O$ and $Q$ operators describe how the evolution of the state $\left|\psi(t,z^{\ast}_t,w^{\ast}_t)\right\rangle$ at time $t$ is influenced by its dependence on the noises $z^{\ast}_s,w^{\ast}_s$ at earlier times $s$.

		The bath correlation functions can be modeled as exponential-decay Ornstein-Uhlenbeck form \cite{finite}, i.e.,
		
		\begin{eqnarray}
			\alpha(t,s)&=&\frac{\Gamma\gamma}{2}(T+\omega_{0}-i\gamma)e^{-(i\omega_{0}+\gamma)\left|t-s\right|},
			\label{eq8}\\
			\eta(t,s)&=&\frac{\Gamma T\gamma}{2}e^{-(-i\omega_{0}+\gamma)\left|t-s\right|}.
			\label{eq9}
		\end{eqnarray}

		From this correlation function, the Markovian to non-Markovian transition can be easily tuned by  parameter $\gamma$. Experimentally this can be realized by changing the tilt angle of the Fabry–Pérot cavity \cite{BLiu}. The reduced density operator of the system can be recovered by calculating the ensemble average of the operator $P_t$	
		
		\begin{equation}
			\rho_{s}=M\left[P_t\right],
		\end{equation}
		where $P_t=\left|\psi(t,z^{\ast}_t,w^{\ast}_t)\right\rangle \left\langle\psi(t,z^{\ast}_t,w^{\ast}_t)\right|$, and $ M[\cdot]$ is the ensemble average over the complex Gaussian stochastic process $z$ or $w$, and $	M[\mathcal{F}]= \prod \limits_{k} \frac{1}{\pi}\int e^{-\left| z \right|^{2}}\mathcal{F}d^{2}z$. Then taking the time-derivative to $\rho_{s}$, we obtain the non-Markovian master equation \cite{yu2004,Gisin}

		\begin{eqnarray}
			\frac{\partial}{\partial t}\rho_{s}&=&-i\left[H_{s},\rho_{s}\right]+[L,M[P_{t}\overline{O}^{\dagger}(t,z^{\ast}_t,w^{\ast}_t)]]-[L^\dagger,M[\overline{O}(t,z^{\ast}_t,w^{\ast}_t)P_{t}]]\nonumber\\&\;&
			+[L^\dagger,M[P_{t}\overline{Q}^{\dagger}(t,z^{\ast}_t,w^{\ast}_t)]]-[L,M[\overline{Q}(t,z^{\ast}_t,w^{\ast}_t)P_{t}]].
		\end{eqnarray}

		Under the weak coupling assumption, the noise dependent $\overline{O}(t,z^{\ast}_t,w^{\ast}_t)$,  $\overline{Q}(t,z^{\ast}_t,w^{\ast}_t)$ operators can be approximated well by noise independent $\overline{O}(t)$,  $\overline{Q}(t)$ operators. Then non-Markovian master equation (Eq.\ref{eq4}) is obtained.

		For the two mode squeezed case, the environmental correlation functions can be written as \cite{Link, Arapat zero},
		
		\begin{equation}
			\alpha_{1}(t,s)=\frac{\Gamma\gamma}{2}(T+\omega_{0}-i\gamma)\left(u^{2}-vue^{-2iw_{0}s}\right)e^{-(i\omega_{0}+\gamma)\left|t-s\right|},
			\label{eq13}
		\end{equation}
		\begin{equation}
			\alpha_{2}(t,s)=\frac{\Gamma\gamma}{2}(T-\omega_{0}-i\gamma)\left(\left|v\right|^{2}-v^{*}ue^{2iw_{0}s}\right)e^{-(-i\omega_{0}+\gamma)\left|t-s\right|},
		\end{equation}
		
		\begin{equation}
			\eta_{1}(t,s)=\frac{\Gamma T\gamma}{2}\left(u^{2}-vue^{2iw_{0}s}\right)e^{-(-i\omega_{0}+\gamma)\left|t-s\right|},
		\end{equation}
		
		\begin{equation}
			\eta_{2}(t,s)=\frac{\Gamma T\gamma}{2}\left(\left|v\right|^{2}-v^{*}ue^{-2iw_{0}s}\right)e^{-(i\omega_{0}+\gamma)\left|t-s\right|},
			\label{eq16}
		\end{equation}
		where $\alpha(t,s)=\alpha_{1}(t,s)+\alpha_{2}(t,s)$, $\eta(t,s)=\eta_{1}(t,s)+\eta_{2}(t,s)$, and $u=cosh(r)$, $v=w\xi$, $w=sinh(r)$. They satisfy the relation $u^{2}-\left|v\right|^{2}=1$. $r=0$ corresponds to EIVS with $\alpha_{2}(t,s)=\eta_{2}(t,s)=0$. When $\theta=0 (\pi)$ the bath squeezes the $p (x)$-quadrature. In this case the $\overline{O}_{1(2)}(t)$, $\overline{Q}_{1(2)}(t)$ operators satisfy

		\begin{eqnarray}
			\frac{\partial}{\partial t}\overline{O}_{1}(t)&=& \alpha_{1}(0,0)L-(iw_{0}+\gamma)\overline{O}_{1}(t)
			-[iH_{s}+L^\dagger\overline{O}_{1}(t)+L^\dagger\overline{O}_{2}(t)
			+L\overline{Q}_{1}(t)+L\overline{Q}_{2}(t),\overline{O}_{1}(t)],
		\end{eqnarray}
		
		\begin{eqnarray}
			\frac{\partial}{\partial t}\overline{O}_{2}(t)&=& \alpha_{2}(0,0)L-(-iw_{0}+\gamma)\overline{O}_{2}(t)
			-[iH_{s}+L^\dagger\overline{O}_{1}(t)+L^\dagger\overline{O}_{2}(t)
			+L\overline{Q}_{1}(t)+L\overline{Q}_{2}(t),\overline{O}_{2}(t)],
		\end{eqnarray}
		
		\begin{eqnarray}
			\frac{\partial}{\partial t}\overline{Q}_{1}(t)&=& \eta_{1}(0,0)L^\dagger-(-iw_{0}+\gamma)\overline{Q}_{1}(t)
			-[iH_{s}+L^\dagger\overline{O}_{1}(t)+L^\dagger\overline{O}_{2}(t)
			+L\overline{Q}_{1}(t)+L\overline{Q}_{2}(t),\overline{Q}_{1}(t)],
		\end{eqnarray}
		
		\begin{eqnarray}
			\frac{\partial}{\partial t}\overline{Q}_{2}(t)&=& \eta_{2}(0,0)L^\dagger-(iw_{0}+\gamma)\overline{Q}_{2}(t)
			-[iH_{s}+L^\dagger\overline{O}_{1}(t)+L^\dagger\overline{O}_{2}(t)
			+L\overline{Q}_{1}(t)+L\overline{Q}_{2}(t),\overline{Q}_{2}(t)],
		\end{eqnarray}
		
		where $\overline{O}_k(t)=\intop_{0}^{t}ds\alpha_k(t,s)O(t,s)$, $\overline{Q}_k(t)=\intop_{0}^{t}ds\beta_k(t,s)Q(t,s)$ ($k=1,2$).

		\section{ B. Steady state solution in Markovian limit}
		
		\setcounter{equation}{0}
		
		\renewcommand\theequation{B.\arabic{equation}}
		
		The Markovian Lindblad master equation can be written as
		
		\begin{eqnarray}
			\frac{\partial}{\partial t}\rho_{s}&=&-i\left[H_{s},\rho_{s}\right]+\frac{\Gamma T}{2}[\left(2L\rho_{s}L^{\dagger}-L^{\dagger}L\rho_{s}-\rho_{s}L^{\dagger}L \right) 
			+\left(2L^{\dagger}\rho_{s}L-LL^{\dagger}\rho_{s}-\rho_{s}LL^{\dagger}\right)],
		\end{eqnarray}

		In the steady state, we have $\frac{\partial}{\partial t}\rho_{s}=0$, i.e., 
		
		\begin{eqnarray}
			-i\left[H_{s},\rho_{s}\right]=\frac{\Gamma T}{2}[\left(2L\rho_{s}L^{\dagger}-L^{\dagger}L\rho_{s}-\rho_{s}L^{\dagger}L \right) 
			+\left(2L^{\dagger}\rho_{s}L-LL^{\dagger}\rho_{s}-\rho_{s}LL^{\dagger}\right)]=0.
		\end{eqnarray}

		Then, we have $\rho_{12}=\rho_{13}=\rho_{14}=\rho_{21}=\rho_{24}=\rho_{31}=\rho_{34}=\rho_{41}=\rho_{42}=\rho_{43}=0$, and

		\begin{equation}
			-2\rho_{11}+\rho_{22}+\rho_{23}+\rho_{32}+\rho_{33}=0,
			\label{eqB19}
		\end{equation}
		
		\begin{equation}
			\rho_{11}-2\rho_{22}-\rho_{23}-\rho_{32}+\rho_{44}=0,
			\label{eqB20}
		\end{equation}
		
		\begin{equation}
			-2i(\omega_1-\omega_2)\rho_{23}+\frac{\Gamma T}{2}(4\rho_{11}-4\rho_{22}-8\rho_{23}-4\rho_{33}+4\rho_{44})=0,
			\label{eqB21}
		\end{equation}
		
		\begin{equation}
			2i(\omega_1-\omega_2)\rho_{32}+\frac{\Gamma T}{2}(4\rho_{11}-4\rho_{22}-8\rho_{32}-4\rho_{33}+4\rho_{44})=0,
			\label{eqB22}
		\end{equation}

		\begin{equation}
			\rho_{11}-\rho_{23}-\rho_{32}-2\rho_{33}+\rho_{44}=0,
			\label{eqB23}
		\end{equation}
		
		\begin{equation}
			\rho_{22}+\rho_{23}+\rho_{32}+\rho_{33}-2\rho_{44}=0.
			\label{eqB24}
		\end{equation}

		When $\omega_1=\omega_2$, we have a solution of $\rho_{23}=\rho_{32}$ and $\rho_{11}=\rho_{44}=2\rho_{22}=2\rho_{33}=2\rho_{23}=2\rho_{32}$. Then using $\rho_{11}+\rho_{22}+\rho_{33}+\rho_{44}=1$, we obtain

		\begin{equation}
			\rho_{11}=\rho_{44}=2\rho_{22}=2\rho_{33}=2\rho_{23}=2\rho_{32}=1/3.
		\end{equation}
		
		The steady state solution is

\begin{equation}
	\left( \begin{array}{cccc}
		\frac{1}{3} & 0 &0 &0 \\
		
		0 & \frac{1}{6} &\frac{1}{6} &0\\
		
		0 & \frac{1}{6} &\frac{1}{6} &0\\
		
		0 & 0 & 0 &\frac{1}{3}
	\end{array} 
	\right).
\end{equation}

		And this is in accordance with our numerical results in Fig.~\ref{fig:1}.

		When $\omega_1\neq\omega_2$, $\rho_{23}=\rho_{32}=0$, we have
		
		\begin{equation}
			-2\rho_{11}+\rho_{22}+\rho_{33}=0,
		\end{equation}
		
		\begin{equation}
			\rho_{11}-2\rho_{22}+\rho_{44}=0,
		\end{equation}
		
		\begin{equation}
			\rho_{11}-2\rho_{33}+\rho_{44}=0,
		\end{equation}
		
		\begin{equation}
			\rho_{22}+\rho_{33}-2\rho_{44}=0,
		\end{equation}
		
		then we have $\rho_{11}=\rho_{22}=\rho_{33}=\rho_{44}=1/4$. The steady state solution is
		
\begin{equation}
	\left( \begin{array}{cccc}
		\frac{1}{4} & 0 &0 &0 \\
		
		0 & \frac{1}{4} &0 &0\\
		
		0 & 0 &\frac{1}{4} &0\\
		
		0 & 0 & 0 &\frac{1}{4}
	\end{array} 
	\right).
\end{equation}
	\end{appendix}
\end{widetext}

In this case the SSQC does not occur.

\end{document}